\documentclass[10pt,letterpaper]{article}

\usepackage[utf8]{inputenc}
\usepackage{hyperref}
\usepackage{lmodern}
\usepackage{amsmath,amsthm,amssymb}
\usepackage{accents}
\usepackage{microtype}
\usepackage[hmargin=3.8cm,vmargin=1.8cm]{geometry}
\usepackage{xcolor}
\usepackage{nicefrac}
\usepackage{bbm}
\usepackage{aliascnt}
\usepackage{mathtools}
\usepackage{sectsty}
\usepackage{tikz}
\usepackage{caption}
\usepackage{float}

\definecolor{Blue}{RGB}{5,55,77}
\definecolor{Red}{RGB}{20,50,10}
\definecolor{purple}{HTML}{8c3fcf}
\hypersetup{
    colorlinks,
    citecolor=Blue,
    filecolor=Blue,
    linkcolor=Blue,
    urlcolor=Blue,
    breaklinks=false,
}

\sectionfont{\large}
\subsectionfont{\normalsize}

\newcommand{\sI}{\mathrm{I}_p}
\newcommand{\qI}{\bar{\mathrm{I}}_p}
\newcommand{\qJ}{\mathrm{J}_p}
\newcommand{\Q}{Q(C)}

\newcommand{\qent}[1]{Q{\mkern-2mu}\left(#1\right)}



\newcommand{\nparen}[1]{\left(#1\right)}   %normal parantheses
\newcommand{\sparen}[1]{\left[#1\right]}   %square parantheses
\newcommand{\lebd}{\,\mathrm{d}}

\newcommand{\cube}{\{0,1\}^n}

\newcommand{\emid}{\,|\,}

\newcommand{\E}{\mathop{{}\mathbb{E}}}

\newcommand{\muti}[2]
    {{\nonscript\mkern2mu}\mathbf{I}{\nonscript\mkern-2mu}\left(#1{\nonscript\mkern-1mu}\mathbin{:}{\nonscript\mkern-1mu}#2\right)}   %mutual info

\newcommand{\kldivb}[2]
    {\mathbf{D}_2{\mkern-2mu}\left(#1\,\middle\|\,#2\right)}

\newcommand{\ent}[1]
    {\mathrm{H}{\nonscript\mkern-2.7mu}\left(#1\right)}

\newcommand{\entb}[1]
    {\mathrm{H}_2{\mkern-2mu}\left(#1\right)}

\newcommand{\tsdiv}[2]
    {\bar{\mathbf{D}}{\mkern-2mu}\left(#1\,\middle\|\,#2\right)}

\newcommand{\tsdivb}[2]
    {\bar{\mathbf{D}}_2{\mkern-2mu}\left(#1\,\middle\|\,#2\right)}

\newcommand{\tsent}[1]
    {\bar{\mathrm{H}}{\nonscript\mkern-2.7mu}\left(#1\right)}

\newcommand{\tsmuti}[2]
    {{\nonscript\mkern2mu}\bar{\mathbf{I}}{\nonscript\mkern-2mu}\left(#1{\nonscript\mkern-1mu}\mathbin{;}{\nonscript\mkern-1mu}#2\right)}

\newtheorem{theorem}{Theorem}[section]

  \newaliascnt{conjecture}{theorem}
  \newtheorem{conjecture}[conjecture]{Conjecture}
  \aliascntresetthe{conjecture}

  \newaliascnt{corollary}{theorem}
  
  \aliascntresetthe{corollary}

  \newaliascnt{definition}{theorem}
  
  \aliascntresetthe{definition}

  \newaliascnt{lemma}{theorem}
  \newtheorem{lemma}[lemma]{Lemma}
  \aliascntresetthe{lemma}

  \newaliascnt{proposition}{theorem}
  
  \aliascntresetthe{proposition}

  \newaliascnt{remark}{theorem}
  
  \aliascntresetthe{remark}

\numberwithin{equation}{section}
\makeatletter
\let\oldtheequation\theequation
\renewcommand\tagform@[1]{\maketag@@@{\ignorespaces#1\unskip\@@italiccorr}}
\renewcommand\theequation{(\oldtheequation)}
\makeatother
\renewcommand{\equationautorefname}{Eq.\kern-2pt}

\usepackage[
  backend=biber,
  style=numeric,
  natbib=true,
  backref,
  url=true,
  maxbibnames=99]{biblatex}
\DefineBibliographyStrings{english}{%
  backrefpage = {cited on page},
  backrefpages = {cited on pages},
}
\begin{document}
\title{\vspace{-30px}
\bf\fontsize{15}{19}\selectfont
  On the most informative Boolean function under noise%
  \vspace{-2px}}
\author{\fontsize{10}{12}\selectfont
\begin{minipage}[t]{0.45\textwidth}\centering
Dimiter Kramer\\[1pt]
{\fontsize{9}{11}\selectfont DeepInfra\\
dimiter@deepinfra.com}
\end{minipage}\hfill
\begin{minipage}[t]{0.45\textwidth}\centering
Mert Saglam\\[1pt]
{\fontsize{9}{11}\selectfont DeepInfra\\
mert@deepinfra.com}
\end{minipage}}
\date{}
\maketitle
\vspace{-8px}

% !TeX root = courtade-kumar.tex
\renewenvironment{abstract}{\small \list{}{%
    \setlength{\leftmargin}{10mm}%
    \setlength{\rightmargin}{\leftmargin}%
    \listparindent 0em%
  }\item\relax{\bf Abstract.}}{\endlist}

\begin{abstract}
Let $X$ be a uniform random bit string and $Y$ be obtained
by flipping each bit of $X$ independently with probability
$p$.
Let $C$ be a random bit obtained as a function of $Y$. We
show that $$\muti{C}{X}\le \muti{Y_1}{X}$$ for all
$p\in[0,1]$, which resolves a conjecture posed by
Courtade and Kumar in 2013.
\end{abstract}

% !TeX root = courtade-kumar.tex
% !TEX TS-program = pdfLaTeX
\section{Introduction}
\label{sec:introduction}
Let $X$ be distributed uniformly on $\cube$ and let $Y$
be obtained by flipping each coordinate of $X$ with
probability $p$, independently.
In \cite{KumarC2013,CourtadeK2014}, Courtade and Kumar ask:
which bit $C$ computed from $Y$ is most informative about
$X$?
They conjecture that taking $C=Y_i$ for some $i\in[n]$
maximizes the information conveyed about $X$. In particular,
their Conjecture 1 states that 
\begin{equation}
\label{eq:ck}\tag{(CK)}
\muti{C}{X}\le\muti{Y_1}{X} = 1 - \entb{p}.
\end{equation}
They further remark that, despite its resemblance to a
simple homework exercise, the statement appears to
involve substantial depth. Indeed, even after extensive
efforts and partial progress
\cite{AnantharamBCJN2017,Arikan2009,BarnesO2020,ChenN2024,
FischerKRSS2004,JavanmardW2026,KindlerOW2015,LiM2021,
OrdentlichSW2015,ParnasRS2002,PichlerMP2016,Samorodnitsky2015,
Samorodnitsky2016,Sarbu2017,Sarbu2017b,YangW2018,Yu2024},
a full resolution of the conjecture remained elusive.

% !TeX root = ../courtade-kumar.tex
\usetikzlibrary{decorations.markings}
% blue/purple scheme
\definecolor{lecol}{RGB}{40,160,235}
\definecolor{gecol}{RGB}{140,82,235}
\def\regionopacity{0.28}
% green/blue scheme (swap in to use; update caption colour names)
% \colorlet{lecol}{green!55!black}
% \colorlet{gecol}{blue!70!black}
% \def\regionopacity{0.18}
\begin{figure}[H]
\centering
\vspace*{0.5em}
\begin{tikzpicture}[scale=2.7,
    corner/.style={font=\small},
    inc/.style={thick,line cap=round,postaction={decorate},
      decoration={markings,mark=at position 0.55 with {\arrow{stealth}}}}]
  % Top edge \sI H -> \qI H has length (\qI - \sI)H; bottom edge has
  % length (\qI - \sI)\bar H. Q sits on the top edge at distance
  % (\qI - \sI)\bar H from \qI H, i.e. directly above \sI \bar H.
  \coordinate (HI)   at (0,1);
  \coordinate (HbI)  at (1.6,1);
  \coordinate (bHI)  at (0.6,0);
  \coordinate (bHbI) at (1.6,0);
  \coordinate (Q)    at (0.6,1);
  % shaded regions; their overlap blends to a third colour
  \begin{scope}
    \clip[rounded corners=6pt] (-0.72,-0.3) rectangle (2.5,1.38);
    % values guaranteed <= I
    \fill[lecol,opacity=\regionopacity]
      (-0.72,-0.3) -- (0.95,-0.3)
      .. controls (1.2,0.35) and (1.0,0.95) .. (1.14,1.38)
      -- (-0.72,1.38) -- cycle;
    \draw[lecol!85,line width=1pt]
      (0.95,-0.3) .. controls (1.2,0.35) and (1.0,0.95) .. (1.14,1.38);
    % values guaranteed >= I(B:X)
    \fill[gecol,opacity=\regionopacity]
      (0.14,1.38) .. controls (0.22,0.65) and (1.3,0.45) .. (2.5,0.38)
      -- (2.5,1.38) -- cycle;
    \draw[gecol!85,line width=1pt]
      (0.14,1.38) .. controls (0.22,0.65) and (1.3,0.45) .. (2.5,0.38);
  \end{scope}
  \node[black!80,font=\small,anchor=south west] at (-0.69,-0.27)
    {$\le\muti{Y_1}{X}$};
  \node[black!80,font=\small,anchor=north east] at (2.47,0.70)
    {$\ge\muti{C}{X}$};
  % edges point in the increasing direction
  \draw[inc] (HI)   -- (HbI);
  \draw[inc] (bHI)  -- (bHbI);
  \draw[inc] (bHI)  -- (HI);
  \draw[inc] (bHbI) -- (HbI);
  \draw[densely dashed,gray!70,shorten >=2pt,shorten <=2pt] (Q) -- (bHI);
  \foreach \p in {HI,HbI,bHI,bHbI} \fill (\p) circle (1.1pt);
  \filldraw[fill=white,thick] (Q) circle (1.3pt);
  % mid anchors are baseline-relative, so the two top labels share a baseline
  \node[corner,anchor=south east,inner sep=4pt] at (HI)   {$\sI\mathrm H$};
  \node[corner,anchor=south west,inner sep=4pt] at (HbI)  {$\qI\mathrm H$};
  \node[corner,anchor=north east,inner sep=4pt] at (bHI)  {$\sI\bar{\mathrm H}$};
  \node[corner,anchor=north west,inner sep=4pt] at (bHbI) {$\qI\bar{\mathrm H}$};
  \node[anchor=south,inner sep=5pt] at (Q) {$\Q$};
\end{tikzpicture}
\captionsetup{font=small,name=Figure,labelformat=default,labelsep=period,labelfont=bf}
\caption{Arrows point toward the larger quantity.
$\Q$ lies on the top edge, and its distance to
$\qI\ent{C}$ equals the length of the bottom edge,
$(\qI-\sI)\,\tsent{C}$.
Blue: quantities that are always at most $\muti{Y_1}{X}$.
Purple: quantities that upper bound $\muti{C}{X}$.
Only $\Q$ lies in both.}
\label{fig:q-square}
\end{figure}
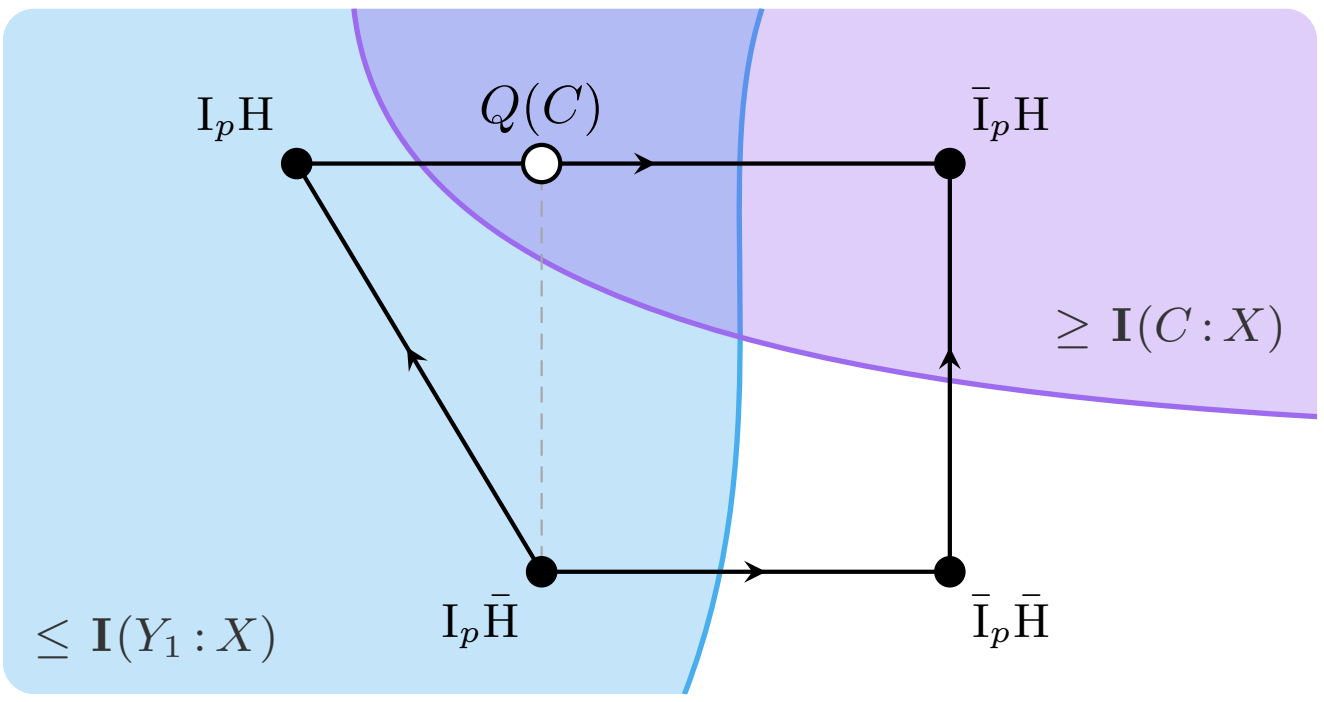

In this work, we observe that if we study the inequality
parametrized by $\mu=\E[C]$, a delicate refinement
of \eqref{eq:ck} interpolating between Shannon and Tsallis
entropies emerges. This refinement admits a proof by induction
on the number of coordinates of $X$, $Y\!$.
Denoting the Tsallis entropy of $C$ by $\tsent{C} = 4\mu(1-\mu)$
and setting $\sI = \muti{Y_1}{X}$ and $\qI = (1-2p)^2$, we define
the progressive entropy of $C$ as
\begin{equation}
\label{eq:Q}\tag{(Q)}
  \Q=\qI\ent{C}-(\qI-\sI)\tsent{C}.
\end{equation}
With this definition in hand, we prove the following.
\begin{theorem}
\label{thm:ck+}
Let $X$ be a uniform random bit string and $Y$ be obtained
by flipping each bit of $X$ with probability $p$,
independently. For a bit $C$ obtained from $Y\!$, we have
\begin{equation*}
\muti{C}{X}\le\qent{C}\le \muti{Y_1}{X}
\end{equation*}
for any $p\in[0,1]$.
\end{theorem}

\subsection{Comparison with earlier bounds}
\label{sec:earlier}
The strong data-processing inequality \cite{AhlswedeG1976}
gives $\muti{C}{X}\le\qI\ent{C}$,
the upper-right corner of \autoref{fig:q-square}.
Courtade and Kumar's \cite{CourtadeK2014} Conjecture~2
concerns the stronger upper bound $\sI\ent{C}$ at the upper-left
corner. This bound was later shown to be too strong in
general \cite{ChandarT2014,KindlerOW2015}.
Our refinement $\qent{C}$ subtracts a Tsallis correction
$(\qI-\sI)\,\tsent{C}$ from the strong data-processing bound
and sits between the two.

\subsection{Concurrent work}
\label{sec:concurrent}
Concurrently with our work, Ky and Tran~\cite{KyT2026},
Chen et al.~\cite{ChenGJL+2026}, and Mahdavifar and
Beirami~\cite{MahdavifarB2026} announced proofs of the
Courtade--Kumar conjecture. Their preprints were first
submitted to arXiv on September~21--22, 2026.
We had obtained a proof using Fourier analysis by
September~17, 2026, and completed writing down the simpler
induction proof, which we present here, by September~20, 2026.
Both versions are recorded in our public repository.\footnote{%
The Fourier proof:
\url{https://github.com/deepinfra/courtade-kumar/commit/1b7f244}.
The induction proof:
\url{https://github.com/deepinfra/courtade-kumar/commit/a07761c}.}

OpenAI subsequently released two further proofs in its
\href{https://github.com/openai/math}{\texttt{openai/math}} repository on October~6,
2026.\footnote{\url{https://openai.com/index/sharing-ai-progress-in-mathematics/}.}
Both manuscripts are dated September~24, 2026.
The first~\cite{OpenAI2026a} derives \eqref{eq:ck} from a
Hellinger-contraction inequality and uses computer-assisted
verification of scalar inequalities.
The second~\cite{OpenAI2026b} proves sharp information
contraction for randomized binary outputs.

\subsection{The continuous case}
\label{sec:continuous}
We note that $\muti{C}{X}\le \Q$ is a surprisingly sharp
bound, which is already violated by a natural
generalization of the problem.
Let $X_\circ$ be distributed uniformly on the unit cube
$[0,1]^n$ and $Y_\circ$ be obtained by independently
resampling each coordinate of $X_\circ$ uniformly from
$[0,1]$ with probability $2p$.
Assume $C_\circ$ is a random bit obtained as a
function of $Y_\circ$ with $\mu=\E\sparen{C_\circ}$.

Let $T_\mu$ be the indicator of
$Y_{\circ,1}\le \mu$, so that $\E\sparen{T_\mu}=\mu$. In
\autoref{lem:ck+vscontinuous} we show that
$\qent{C_\circ} < \muti{T_\mu}{X_\circ}$ for $0<\mu<1$,
$\mu\ne1/2$, and $0<p<1/2$, which shows that $\qent{C_\circ}$
is too strong for the continuous problem. We conjecture that
the threshold functions maximize mutual information at
each fixed mean.
\begin{conjecture}
\label{conj:continuous}
Let $X_\circ$, $Y_\circ$ and $C_\circ$ be as constructed
and $T_\mu$ be the indicator of the event $Y_{\circ,1}\le\mu$
as above. It holds that
\begin{equation*}
  \muti{C_\circ}{X_\circ}\le\muti{T_\mu}{X_\circ}
\end{equation*}
for $0\le p\le1/2$ and $0\le\mu\le1$.
\end{conjecture}

Note that a proof of \autoref{conj:continuous} directly
implies \eqref{eq:ck}---though not our
\autoref{thm:ck+}---since given a strategy to output $C$
from $Y$ revealing $\muti{C}{X}$ bits of information, we can
quantize $Y_\circ$ by rounding each coordinate to $0$ or $1$
with threshold $1/2$ and apply the strategy to obtain
$C_\circ$. We get $\muti{C_\circ}{X_\circ}=\muti{C}{X}$
and it can be shown that
$\muti{T_\mu}{X_\circ}\le\muti{Y_1}{X}$.
\section{Preliminaries}
\label{sec:preliminaries}
We use $\ent{\cdot}$ and $\muti{\cdot}{\cdot}$ for Shannon
entropy and mutual information, with logarithms in base $2$.
We denote the natural logarithm by $\ln$. We only need
Tsallis entropy for bits, while the observed variables may
be arbitrary. For a bit $C$ with mean $\mu$, we define
\begin{equation*}
  \tsent{C}=4\mu(1-\mu).
\end{equation*}
For a bit $C$ and an arbitrary random variable $Z$,
Tsallis information is defined by
\begin{equation*}
  \tsmuti{C}{Z}=\tsent{C}-\tsent{C\emid Z}.
\end{equation*}
Note that for bits $B$, $C$, the quantities
$\tsmuti{B}{C}$ and $\tsmuti{C}{B}$ need
not be equal. We denote
\begin{equation*}
\begin{aligned}
  \sI=1-\entb{p},\qquad \qI&=(1-2p)^2,\\
  \text{and}\qquad \qJ&=\qI-\sI.
\end{aligned}
\end{equation*}
Thus $\sI=\muti{Y_i}{X_i}$ and $\qI=\tsmuti{Y_i}{X_i}$.
With these shorthands, we have
\begin{equation*}
  \qent{C}=\qI\ent{C}-\qJ\tsent{C}.
\end{equation*}
For all our entropy and information quantities
$\ent{\cdot}$, $\muti{\cdot}{\cdot}$, $\tsent{\cdot}$,
$\tsmuti{\cdot}{\cdot}$, and $Q$, conditioning on $U$ means
evaluating the quantity under the conditional law given $U$
and taking expectation over $U$, with $p$ held fixed.
For instance,
$\qent{C\emid U} = \E_{u\sim U}[\qent{C\emid U=u}]$.
Next, we gather some facts about Tsallis and Shannon
information.
\begin{lemma}
\label{lem:tsallis-properties}
Let $B$ and $C$ be bits and $Z$, $W$, $U$ be arbitrary
random variables. We have
\leavevmode
\begin{enumerate}
\setlength{\itemsep}{4pt}
\item\label{item:tsallis-bounds}
$0\le\tsmuti{C}{Z\emid U}\le\tsent{C\emid U}$ and
$\tsent{C\emid U}\le\ent{C\emid U}
\le\bigl(1+\tsent{C\emid U}\bigr)/2$.
\item\label{item:tsallis-chain}
$\tsmuti{C}{Z,W\emid U}
=\tsmuti{C}{Z\emid U}+\tsmuti{C}{W\emid Z,U}$.
If $C-Z-W$ is a Markov chain, then
$\tsmuti{C}{W}\le\tsmuti{C}{Z}$.
\item\label{item:tsallis-shannon}
$\tsmuti{C}{Z\emid U}\le2\ln2\,\muti{C}{Z\emid U}$.
If $C$ is unbiased and independent of $U$, then also
$\muti{C}{Z\emid U}\le\tsmuti{C}{Z\emid U}$.
\item\label{item:tsallis-bits}
$\tsent{C}\tsmuti{B}{C}=\tsent{B}\tsmuti{C}{B}$.
If $B$ is unbiased, then
$\tsmuti{C}{B}=\bigl(2\Pr[B=C]-1\bigr)^2$.
\item\label{item:tsallis-conditioning}
If $B$ is independent of $U$, then
$\tsmuti{B}{C\emid U}\,\tsent{C\emid U}
\ge\tsent{B}\,\tsmuti{C}{B}$.
\item\label{item:tsallis-noise}
$0\le\qJ\le\sI$ for every $p\in[0,1]$.
\end{enumerate}
\end{lemma}

Proofs of the preceding and the next lemma are given in
\autoref{sec:quadi}. Let us define
\begin{equation*}
  c(\mu)=\frac{\ln(1/\mu)}{4(1-\mu)^2}
  \qquad\nparen{1/2\le\mu<1},
\end{equation*}
extending it to $0<\mu<1/2$ by $c(\mu)=c(1-\mu)$.

\begin{lemma}
\label{lem:biased-comparison}
For a nonconstant bit $C$ with mean $\mu$ and every random variable $U$,
\begin{equation*}
  \muti{C}{U}\le\frac{c(\mu)}{\ln2}\tsmuti{C}{U},
  \qquad\text{and}\qquad
  \tsent{C}\,c(\mu)\le1.
\end{equation*}
\end{lemma}

% !TeX root = courtade-kumar.tex
% !TEX TS-program = pdfLaTeX

\section{The proof}
\label{sec:proof}
Our proof proceeds by induction on $n$, the number of
coordinates of $X,Y$. Given the random bit $C$ and the
bit string $Y\!$, we adaptively pick a coordinate $i\in[n]$ and
establish the bound
\begin{align}
  \tag{($\star$)}\label{eq:star}
  \qent{C} - \muti{C}{X} \ge \frac{1}{1+ \lambda}
  \nparen{\qent{C\emid Y_i}-\muti{C}{X\emid Y_i}}
\end{align}
for some $\lambda\ge 0$.
When $Y_i$ is fixed, $C$ depends on $n-1$ coordinates and
therefore the induction hypothesis applies for $Y_i=0$ and
$Y_i=1$ separately, showing that
$\qent{C\emid Y_i}\ge \muti{C}{X\emid Y_i}$.
Plugging into \eqref{eq:star}, we get
$\qent{C}\ge\muti{C}{X}$, which implies \eqref{eq:ck}
as we show next.
\begin{lemma}
\label{lem:ck+to_ck}
For all $p\in[0,1]$, we have
$\qent{C}\le \muti{Y_1}{X}$.
\end{lemma}
\begin{proof}
From \hyperref[item:tsallis-noise]{Lemma~\ref*{lem:tsallis-properties}(\ref*{item:tsallis-noise})}
and \hyperref[item:tsallis-bounds]{(\ref*{item:tsallis-bounds})},
we have $\qJ\le\sI$ and
$\ent{C}-\tsent{C}\le1-\ent{C}$. Let us write
\begin{align*}
  \qent{C} - \sI\ent{C}
  &= \qJ\ent{C} - \qJ\tsent{C}\\
  &=   \qJ\nparen{\ent{C}-\tsent{C}}\\
  &\le \sI \nparen{1-\ent{C}}.
\end{align*}
Therefore
$\Q\le \sI\nparen{1-\ent{C}} + \sI\,\ent{C} = \muti{Y_1}{X}$
as desired.
\end{proof}
The property we want from the coordinate $i$ is fairly mild:
$\tsmuti{C}{Y_i}\le 1/4$. The next lemma shows that such a
coordinate always exists.

\begin{lemma}
\label{lem:good-coordinate}
When $n\ge2$, there exists an $i\in[n]$ such that
$\tsmuti{C}{Y_i}\le 1/4$.
\end{lemma}
\begin{proof}
Let $i,j\in[n]$ be distinct. We may relabel $Y_i,Y_j$ so that
$\E[C(2Y_i-1)]$ and $\E[C(2Y_j-1)]$ are nonnegative. Since
$Y_i,Y_j$ are independent uniform bits,
\begin{equation*}
  \sqrt{\tsmuti{C}{Y_i}}+\sqrt{\tsmuti{C}{Y_j}}
  =2\E\sparen{C(2Y_i-1)+C(2Y_j-1)}
  \le4\Pr[Y_i=Y_j=1]=1,
\end{equation*}
using $0\le C\le1$. At least one of the two square roots
is at most $1/2$.
\end{proof}
Let us fix such an $i$ satisfying $\tsmuti{C}{Y_i}\le1/4$
and use the shorthands $A=X_i$, $B=Y_i$, and $U= X_{-i}$
in the rest of the paper.
The following lemma is the crux of our argument,
which establishes \eqref{eq:star} giving the induction step.
\begin{lemma}
\label{lem:ind-step}
Assuming $\tsmuti{C}{B}\le1/4$, we have
\begin{align*}
  \qent{C} - \muti{C}{X} \ge \frac{1}{1+ \lambda}
  \nparen{\qent{C\emid B}-\muti{C}{X\emid B}}
\end{align*}
for some $\lambda\ge 0$.
\end{lemma}
\autoref{sec:ind-step} is devoted to proving \autoref{lem:ind-step};
here we show that it implies the main theorem.
\begingroup
\newtheorem*{restatedtheorem}{Theorem~\ref{thm:ck+}}
\begin{restatedtheorem}[restated]
It holds that $\muti{C}{X}\le \qent{C}\le\muti{Y_1}{X}$.
\end{restatedtheorem}
\endgroup
\begin{proof}
The second inequality is \autoref{lem:ck+to_ck}.
For the first, the cases $p\in\{0,1/2,1\}$ and constant $C$
are immediate. Complementing $Y$ if necessary lets us assume
$0<p<1/2$. We proceed by induction on $n$.
When $n=1$, the bit $C$ is a constant, $Y_1$, or its negation,
and the bound holds with equality.
For $n\ge2$, by \autoref{lem:good-coordinate} there exists a coordinate
$i$ with $\tsmuti{C}{Y_i}\le1/4$.
Conditioning on $B=Y_i$ reduces the problem to $n-1$ coordinates,
so the induction hypothesis applies to the $B=0$ and $B=1$ cases
individually, giving
\begin{equation*}
  \muti{C}{X\emid B}=\muti{C}{U\emid B}
  \le\qent{C\emid B}.
\end{equation*}
Applying \autoref{lem:ind-step}, we conclude
$\qent{C}-\muti{C}{X}\ge0$.
\end{proof}

\subsection{A comparison lemma}
\label{sec:comparison}

Let $U$ be a random variable and $A,B$ be unbiased bits jointly
independent of $U$. Let $C$ be a bit obtained from $(B,U)$ using
private randomness, with $\tsent{C\emid U}>0$.
We compare $C$ with an unbiased bit $C'$ obtained from $B$ alone
using private randomness, subject to the matching condition
\begin{equation}
  \tag{(B)}
  \label{eq:comparison-match}
  \muti{B}{C'}=
  \frac{\muti{B}{C\emid U}}{\tsent{C\emid U}}.
\end{equation}

\begin{lemma}
\label{lem:comparison}
A bit $C'$ satisfying \eqref{eq:comparison-match} exists.
Every such bit satisfies
\begin{equation*}
  \muti{A}{C'}\ge
  \frac{\muti{A}{C\emid U}}{\tsent{C\emid U}}.
\end{equation*}
\end{lemma}
\begin{proof}
For the fixed channel between $A$ and $B$, let $F$ be the
function determined by
\begin{equation*}
F\bigl(\muti{B}{D}\bigr)=\muti{A}{D}
\end{equation*}
as $D$ ranges over unbiased bits generated from $B$.
It can be shown that $F$ is increasing and
concave (cf.\ \autoref{lem:comparison-calculus}), and
\begin{align*}
  \muti{B}{C\emid U=u}&\le\tsent{C\emid U=u},\\
  \muti{A}{C\emid U=u}
  &\le\tsent{C\emid U=u}
  F\left(\frac{\muti{B}{C\emid U=u}}
  {\tsent{C\emid U=u}}\right).
\end{align*}
Values of $u$ with zero conditional Tsallis entropy contribute
zero to both informations and may be omitted.
The first inequality, averaged over $U$, places the right-hand
side of \eqref{eq:comparison-match} in $[0,1]$.
Varying the noise in the channel from $B$ to $C'$ therefore
gives a matching bit.
Applying Jensen's inequality to the second inequality, with weights
proportional to $\tsent{C\emid U=u}$, gives
\begin{equation*}
  \muti{A}{C\emid U}
  \le\tsent{C\emid U}
  F\left(\frac{\muti{B}{C\emid U}}{\tsent{C\emid U}}\right)
  =\tsent{C\emid U}\muti{A}{C'},
\end{equation*}
as required.
\end{proof}

\begin{lemma}
\label{lem:comparison-correlation}
Every bit $C'$ satisfying \eqref{eq:comparison-match} also satisfies
\begin{equation*}
  \tsmuti{C'}{B}\ge
  \frac{\tsmuti{C}{B}^{\phantom{2}}}{\tsent{C\emid U}^{2}}.
\end{equation*}
\end{lemma}
\begin{proof}
Let us write $v=\tsent{C\emid U}$ and
$d=\E[(2B-1)(2C-1)]$.
Since $B$ is unbiased, $\tsmuti{C}{B}=d^2$.
As shown in \autoref{app:comparison}, $|d|\le v$ and
\begin{equation*}
  \frac{\muti{B}{C\emid U}}v
  \ge\kldivb{(1+d/v)/2}{1/2}.
\end{equation*}
If $\theta=\E[(2B-1)(2C'-1)]$, then
$\tsmuti{C'}{B}=\theta^2$ and
$\muti{B}{C'}=\kldivb{(1+\theta)/2}{1/2}$.
The matching identity and monotonicity of this divergence in
$|\theta|$ imply $\theta^2\ge d^2/v^2$.
\end{proof}

\subsection{The induction step}
\label{sec:ind-step}
Recalling that $A=X_i$, $B=Y_i$ and $U=X_{-i}$,
in this section we would like to lower bound the gap
$\qent{C} - \muti{C}{X}$ by a factor of the conditional gap
\begin{align*}
\Delta=\qent{C\emid B} - \muti{C}{X\emid B}.
\end{align*}

We obtain two lower bounds on $\qent{C}-\muti{C}{X}$.
The first involves $\Delta$ and benefits from $C$ depending on
$U$, through a positive term in $\tsmuti{C}{U}$.
The second has a negative term in $\tsmuti{C}{U}$ and
benefits from weaker dependence.
We add $\lambda$ times the
second bound to the first, choosing $\lambda$ to cancel these
terms, so that the bound works whether or not $C$ depends on $U$.

To analyze the remaining terms, we introduce an unbiased
comparison bit $C'$ obtained from $B$ using private randomness.
The comparisons in \autoref{lem:comparison} and
\autoref{lem:comparison-correlation}, together with our choice
of coordinate with $\tsmuti{C}{B}\le1/4$, allow us to charge
these terms to quantities
involving $\muti{A}{C'}$ and $\muti{B}{C'}$.
What remains is an inequality about the three unbiased bits in the
Markov chain $A-B-C'$, which we analyze in \autoref{sec:3bit}.

% !TeX root = ../courtade-kumar.tex
\usetikzlibrary{arrows.meta,decorations.pathreplacing}
\begin{figure}[H]
\centering
% Previous palette: originalbit FF8040 (coral), comparisonbit 00BFA5 (teal).
\definecolor{originalbit}{HTML}{F252A0}
\definecolor{comparisonbit}{HTML}{E5A900}
\begin{tikzpicture}[
    cell label/.style={minimum width=0.6cm,minimum height=0.6cm,
      inner sep=0pt,outer sep=0pt},
    dependency/.style={-{Stealth[length=1.65mm,width=1.15mm]},
      line width=0.65pt,draw=black!65,shorten <=2pt,shorten >=2pt},
    group brace/.style={decorate,
      decoration={brace,amplitude=4pt},draw=black!40,line width=0.45pt}]
  % The selected coordinate is placed at the right in both rows.
  \fill[lecol,opacity=0.28] (4.8,-0.3) rectangle (5.4,0.3);
  \fill[gecol,opacity=0.28] (4.8,-1.25) rectangle (5.4,-0.65);
  \foreach \y in {0,-0.95} {
    \draw[black!40,line width=0.5pt]
      (0,\y-0.3) rectangle (5.4,\y+0.3);
  }
  \foreach \k in {1,2,6,7} {
    \draw[black!25,dashed,line width=0.35pt]
      (0.6*\k,-0.3) -- (0.6*\k,0.3);
  }
  \foreach \k in {1,...,8} {
    \draw[black!30,line width=0.35pt]
      (0.6*\k,-1.25) -- (0.6*\k,-0.65);
  }
  \draw[lecol!85,line width=1pt] (4.8,-0.3) rectangle (5.4,0.3);
  \draw[gecol!85,line width=1pt] (4.8,-1.25) rectangle (5.4,-0.65);
  \node[cell label] (B) at (5.1,0) {$B$};
  \node[cell label] (A) at (5.1,-0.95) {$A$};
  \node[anchor=east] at (-0.24,0) {$Y$};
  \node[anchor=east] at (-0.24,-0.95) {$X$};

  \node[cell label,minimum width=4.8cm] (Yrest) at (2.4,0) {$Y_{-i}$};
  \draw[group brace] (4.8,-1.40) -- (0,-1.40);
  \node[anchor=north,inner sep=2pt] at (2.4,-1.57) {$U=X_{-i}$};

  \node[cell label,draw=originalbit!85,fill=originalbit!12,line width=1pt]
    (C) at (5.1,1.65) {$C$};
  \node[cell label,draw=comparisonbit!85,fill=comparisonbit!12,line width=1pt]
    (Cprime) at (6.25,1.65) {$C'$};
  \draw[dependency] (C.south west) to[out=205,in=65] (Yrest.north);
  \draw[dependency] (C.south) -- (B.north);
  \draw[dependency] (Cprime.south west) to[out=245,in=65] (B.north east);
\end{tikzpicture}
\captionsetup{font=small,name=Figure,labelformat=default,labelsep=period,labelfont=bf}
\caption{Arrows indicate dependence. The original bit $C$ depends on
$Y_{-i}$ and $B$, while the comparison bit $C'$ depends only on $B$
and private randomness.}
\label{fig:comparison-bit}
\end{figure}
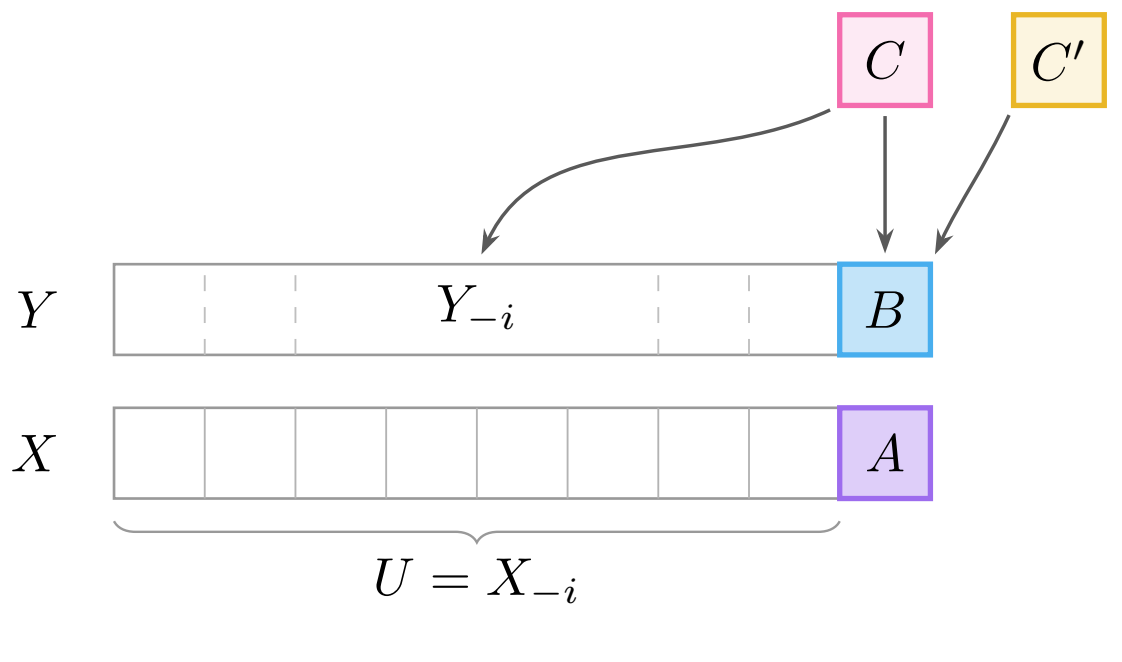

Let us start with a lemma that we use in obtaining the first
lower bound.
\begin{lemma}
\label{lem:conditioning-gain}
If $\tsent{C\emid U}>0$, then
\begin{equation*}
  \muti{B}{C\emid U}-\muti{B}{C}
  \ge
  \frac{1}{2\ln 2}
  \frac{\tsmuti{C}{B}\,\tsmuti{C}{U}}
  {\tsent{C}\,\tsent{C\emid U}}.
\end{equation*}
\end{lemma}
\begin{proof}
Since $B$ and $U$ are independent, the Shannon and Tsallis
chain rules give
\begin{align*}
  \muti{B}{C\emid U}-\muti{B}{C}
  &=\muti{B}{U\emid C},\\
  \tsmuti{B}{C\emid U}-\tsmuti{B}{C}
  &=\tsmuti{B}{U\emid C}.
\end{align*}
Since $B$ is unbiased and independent of $U$, parts
\ref{item:tsallis-shannon}--\ref{item:tsallis-conditioning}
of \autoref{lem:tsallis-properties} give
\begin{align*}
  \muti{B}{U\emid C}
  &\ge\frac{\tsmuti{B}{U\emid C}}{2\ln2}\\
  &\ge\frac{\tsmuti{C}{B}}{2\ln2}
  \left(\frac1{\tsent{C\emid U}}-\frac1{\tsent{C}}\right)\\
  &=\frac{\tsmuti{C}{B}\,\tsmuti{C}{U}}
  {2\ln2\,\tsent{C}\,\tsent{C\emid U}},
\end{align*}
as required.
\end{proof}

When $\tsent{C\emid U}>0$, let $C'$ be an unbiased bit obtained
from $B$ alone using private randomness and satisfying
\eqref{eq:comparison-match}.

\begin{lemma}
\label{lem:gap-first}
If $\tsent{C\emid U}>0$, then
\begin{equation}
  \label{eq:gap-first}
  \begin{aligned}
    \qent{C}-\muti{C}{X}\ge \Delta
    &+\tsent{C\emid U}
    \nparen{\qI\muti{B}{C'}-\muti{A}{C'}}\\
    &+\frac{(1-\qI)\tsmuti{C}{B}\,\tsmuti{C}{U}}
    {2\ln2\,\tsent{C}\,\tsent{C\emid U}}
    -\qJ\tsmuti{C}{B}.
  \end{aligned}
\end{equation}
\end{lemma}
\begin{proof}
The chain rules for Shannon and Tsallis information give
\begin{align*}
  \qent{C}-\muti{C}{X}-\Delta
  =\muti{C}{B\emid X}
  -(1-\qI)\muti{B}{C}-\qJ\tsmuti{C}{B}.
\end{align*}
The chain rule and \autoref{lem:comparison} give
\begin{align*}
  \muti{C}{B\emid X}
  &=\muti{B}{C\emid U}-\muti{A}{C\emid U}\\
  &\ge(1-\qI)\muti{B}{C\emid U}
  +\tsent{C\emid U}
  \bigl[\qI\muti{B}{C'}-\muti{A}{C'}\bigr].
\end{align*}
Substituting this into the preceding identity and applying
\autoref{lem:conditioning-gain} gives \eqref{eq:gap-first}.
\end{proof}

\begin{lemma}
\label{lem:gap-second}
If $\tsent{C\emid U}>0$, then
\begin{align}
  \label{eq:gap-second}
  \qent{C}-\muti{C}{X}
  &\ge\tsent{C\emid U}\muti{A}{B\emid C'}
  -\left[\frac{(1-\qI)c(\mu)}{\ln2}+\qJ\right]\tsmuti{C}{U}.
\end{align}
\end{lemma}
\begin{proof}
Using the comparison in
$\muti{C}{X}=\muti{C}{U}+\muti{A}{C\emid U}$ gives
\begin{align*}
  \qent{C}-\muti{C}{X}
  &\ge\qI\ent{C}-\qJ\tsent{C}
  -\muti{C}{U}-\tsent{C\emid U}\muti{A}{C'}\\
  &=\tsent{C\emid U}\muti{A}{B\emid C'}
  +\qI\bigl[\ent{C\emid U}-\tsent{C\emid U}\bigr]\\
  &\quad-(1-\qI)\muti{C}{U}-\qJ\tsmuti{C}{U}.
\end{align*}
Here we used $\sI=\muti{A}{B}$ and
$\muti{A}{B\emid C'}=\muti{A}{B}-\muti{A}{C'}$.
Now $\ent{C\emid U}\ge\tsent{C\emid U}$ and
\autoref{lem:biased-comparison} give \eqref{eq:gap-second}.
\end{proof}

For $0<p<1$, we use the shorthand
\begin{equation*}
  \alpha_p=\frac{\qJ\ln2}{1-\qI}.
\end{equation*}
The following fact about the unbiased chain $A-B-C'$ is the
remaining ingredient. We state it here and prove it in
\autoref{sec:3bit}.

\begin{lemma}
\label{lem:three-bit}
Let $C'$ be an unbiased bit obtained from $B$ using private
randomness. If $0<\qI<1$ and $\tsmuti{B}{C'}>0$, then
\begin{align*}
  \frac{\sI-\muti{A}{C'}}
  {2(1+\alpha_p)}
  +
  \frac{2\nparen{\qI\muti{B}{C'}-\muti{A}{C'}}}
  {\sqrt{\tsmuti{B}{C'}}}
  \ge\qJ.
\end{align*}
\end{lemma}

\begin{proof}[Proof of \autoref{lem:ind-step}]
The cases $p\in\{0,1/2,1\}$ are immediate, so we may assume
$0<p<1/2$, complementing $Y$ if necessary.
If $\tsmuti{C}{B}=0$, then $C$ and $B$ are independent.
Consequently $\qent{C}=\qent{C\emid B}$ and
$\muti{C}{X\emid B}\ge\muti{C}{X}$, proving the claim with
$\lambda=0$.

We may therefore assume $\tsmuti{C}{B}>0$.
Then $\tsent{C\emid U}>0$:
otherwise $C$ would be determined by $U$ and hence independent of
$B$. By \autoref{lem:comparison}, there is an unbiased bit $C'$
obtained from $B$ alone and satisfying \eqref{eq:comparison-match}.
Let us write
\begin{equation*}
  \lambda=
  \frac{\tsmuti{C}{B}}
  {2\tsent{C}\,\tsent{C\emid U}
  \left(c(\mu)+\alpha_p\right)}.
\end{equation*}
All factors are positive. When we multiply \eqref{eq:gap-second} by
$\lambda$ and add it to \eqref{eq:gap-first},
the terms involving $\tsmuti{C}{U}$ cancel, giving
\begin{equation}
  \label{eq:gap-combined}
  \begin{aligned}
    &(1+\lambda)\bigl[\qent{C}-\muti{C}{X}\bigr]\\
    &\quad\ge\qent{C\emid B}-\muti{C}{X\emid B}
    -\qJ\tsmuti{C}{B}\\
    &\qquad+\tsent{C\emid U}
    \nparen{\qI\muti{B}{C'}-\muti{A}{C'}}
    +\lambda\tsent{C\emid U}\muti{A}{B\emid C'}.
  \end{aligned}
\end{equation}

By \autoref{lem:comparison-correlation} and the choice of coordinate,
\begin{equation*}
  \frac{\tsmuti{C}{B}}{\tsent{C\emid U}}
  \le\sqrt{\tsmuti{C}{B}\,\tsmuti{C'}{B}}
  \le\frac12\sqrt{\tsmuti{C'}{B}}.
\end{equation*}
Also, \autoref{lem:biased-comparison} and $0<\tsent{C}\le1$ give
\begin{equation*}
  \tsent{C}\left(c(\mu)+\alpha_p\right)
  \le1+\tsent{C}\alpha_p
  \le1+\alpha_p.
\end{equation*}
Consequently,
\begin{equation*}
  \frac{\lambda\tsent{C\emid U}}{\tsmuti{C}{B}}
  =\frac1{2\tsent{C}
  \left(c(\mu)+\alpha_p\right)}
  \ge\frac1{2(1+\alpha_p)}.
\end{equation*}
The two information quantities multiplied by these coefficients
are nonnegative. For the first, the binary divergence series gives
$\muti{A}{C'}\le\tsmuti{A}{B}\muti{B}{C'}$; the second
is a conditional mutual information.
Applying \autoref{lem:three-bit}, we therefore obtain
\begin{align*}
  &\frac{\tsent{C\emid U}}{\tsmuti{C}{B}}
  \nparen{\qI\muti{B}{C'}-\muti{A}{C'}}
  +\frac{\lambda\tsent{C\emid U}}{\tsmuti{C}{B}}
  \muti{A}{B\emid C'}\ge\qJ.
\end{align*}
Thus the last three terms in \eqref{eq:gap-combined} have
nonnegative sum. Dividing by $1+\lambda$ proves the lemma.
\end{proof}

\subsection{The three-bit lemma}
\label{sec:3bit}

In this subsection, let $A-B-C$ be an unbiased
binary Markov chain with $0<\tsmuti{A}{B}<1$ and
$\tsmuti{B}{C}>0$.
% !TeX root = courtade-kumar.tex
% Included in the three-bit subsection of proof.tex.
For the calculation, let us write $\rho$ and $u$ for the absolute
correlations of $A,B$ and $B,C$, respectively, and let
\begin{equation*}
  f(t)=\kldivb{(1+t)/2}{1/2}.
\end{equation*}
Thus
\begin{equation*}
  \muti{A}{B}=f(\rho),\quad
  \muti{B}{C}=f(u),\quad
  \muti{A}{C}=f(\rho u),\quad
  \muti{A}{B\emid C}=f(\rho)-f(\rho u).
\end{equation*}
Let us also write
\begin{equation*}
  r=\rho^2,\qquad
  \alpha=\frac{(\rho^2-f(\rho))\ln2}{1-\rho^2}.
\end{equation*}
In these parameters, it suffices to prove
\begin{equation}
  \label{eq:three-bit-scalar-target}
  \frac{f(\rho)-f(\rho u)}{2(1+\alpha)}
  +\frac{2\bigl(\rho^2 f(u)-f(\rho u)\bigr)}u
  >\rho^2-f(\rho).
\end{equation}
We will prove this by expanding in $r$ and showing that every
coefficient is positive. For integers $N\ge1$, let us define the divergence tails
\begin{align*}
  b_N&=\sum_{k>N}\frac{1}{2k(2k-1)}
  =\int_0^1\frac{x^{2N}}{1+x}\,\mathrm{d}x,
  &s_N(u)&=\sum_{k>N}\frac{u^{2k}}{2k(2k-1)}
  =\int_0^u\frac{(u-x)x^{2N}}{1-x^2}\,\mathrm{d}x.
\end{align*}
Let us set $b_0=1$ as a separate convention, and define
\begin{align*}
  T_N&=\sum_{i=0}^{N-1}b_i,
  &B_N&=\sum_{i=0}^{N-1}b_i b_{N-i},\\
  G_N(u)&=\sum_{i=0}^{N-1}b_i s_{N-i}(u),
  &C_N(u)&=\sum_{k=1}^N\frac{1-u^{2k}}{2k(2k-1)}.
\end{align*}
The first ingredient compares the divergence tails in two ways.
\begin{lemma}
\label{lem:scalar-tail-order}
For $0<u<1$ and $N\ge1$,
\begin{equation}
  \label{eq:scalar-convolution-order}
  G_N(u)\ge B_N\frac{s_N(u)}{b_N}.
\end{equation}
For $N\ge2$, we also have
\begin{equation}
  \label{eq:scalar-rescaled-order}
  \frac{s_N(u)}{u b_N}\ge\frac{s_2(u^{N/2})}{u^{N/2} b_2}.
\end{equation}
\end{lemma}
\begin{proof}
First, $s_j(u)/b_j$ is the expectation of the decreasing function
$(u-x)_+/(1-x)$ under the probability density proportional to
$x^{2j}/(1+x)$ on $(0,1)$. Increasing $j$ tilts this density by an
increasing function, so $s_j(u)/b_j$ decreases with $j$.
For $1\le j\le N$, therefore,
$s_j(u)\ge b_j s_N(u)/b_N$; summing proves
\eqref{eq:scalar-convolution-order}.
Here is the elementary order principle used in this argument.
If $Z,Z'$ are independent with the same distribution, then
\begin{equation*}
  \operatorname{Cov}(f(Z),g(Z))
  =\frac12\E\sparen{(f(Z)-f(Z'))(g(Z)-g(Z'))}.
\end{equation*}
It is nonnegative when both functions increase and nonpositive when
one increases and the other decreases. Applying this identity to a
density ratio gives the stated comparison of expectations.
For the second comparison, let us fix $a\ge0$ and write
\begin{equation*}
  M_N(a)=\frac{s_N(e^{-a/N})}{e^{-a/N}b_N}.
\end{equation*}
Changing variables in the two tail integrals gives
\begin{equation}
  \label{eq:scalar-tail-kernel}
  M_N(a)=e^{-2a}\E_{\pi_N}
  \sparen{\frac{\sinh(z/N)}{\sinh((z+a)/N)}},
  \qquad
  \pi_N(\mathrm{d}z)\ \propto
  \frac{e^{-2z}}{1+e^{z/N}}\,\mathrm{d}z,
  \quad z>0.
\end{equation}
The quotient inside the expectation increases with $z$ because
$\coth$ decreases. It also increases with $N$, since
$x\coth x$ increases for $x>0$:
its derivative has the sign of $\sinh x\cosh x-x>0$.
Moreover, for $K>N$ the density ratio $\pi_K/\pi_N$ increases with
$z$, as its logarithmic derivative is
\begin{equation*}
  \frac{N^{-1}}{1+e^{-z/N}}
  -\frac{K^{-1}}{1+e^{-z/K}}>0.
\end{equation*}
The order principle and then the pointwise comparison of quotients
show that $M_N(a)$ increases with $N$.
Taking $a=-N\ln u$ gives
$e^{-a/N}=u$ and $e^{-a/2}=u^{N/2}$, proving
\eqref{eq:scalar-rescaled-order}.
All the densities are integrable; at $a=0$ the quotient in
\eqref{eq:scalar-tail-kernel} equals one.
\end{proof}
The second ingredient gives estimates uniform in the coefficient
index $N$. Let us write
\begin{equation*}
  e_N=\frac12\sum_{k=1}^N\frac{1}{2k-1},\qquad
  R_N=2T_N-1,\qquad x=1-u^{2N}.
\end{equation*}

\begin{lemma}
\label{lem:scalar-moments}
For $N\ge2$ and $0\le u\le1$,
\begin{equation*}
  B_N\le b_NR_N,\qquad
  \frac{R_N}{2e_N}<\frac{19}{18},\qquad
  \frac{C_N(u)}{2e_Nb_N}
  \ge 2x+\frac34x^2+\frac18x^3.
\end{equation*}
\end{lemma}
\begin{proof}
The tail integral and finite summation give
\begin{equation*}
  \frac{1}{4N+2}\le b_N\le\frac{1}{4N+1},\qquad
  T_N=1-\ln 2+e_N+Nb_N.
\end{equation*}
Indeed, the integral bounds follow from
$1+x\le2$ and $1+x\ge2\sqrt{x}$.
For positive integers $i,j$, they imply
\begin{equation*}
  b_i^{-1}+b_j^{-1}\ge4(i+j)+2\ge b_{i+j}^{-1},
  \qquad
  b_i b_j\le b_{i+j}(b_i+b_j).
\end{equation*}
Summing over $i+j=N$ and including the $b_0b_N=b_N$ term proves
$B_N\le b_NR_N$.
For the second bound, let us write $t=\sqrt{2/N}\in(0,1]$.
We will use
\begin{equation}
  \label{eq:scalar-harmonic-bound}
  e_N\ge\frac76-\frac t2\ge1-\frac t3.
\end{equation}
The first inequality is equality at $N=2$.
Its right-hand side increases from $N$ to $N+1$ by at most
$e_{N+1}-e_N=1/(4N+2)$, because
\begin{equation*}
  \sqrt{N}\sqrt{N+1}(\sqrt{N}+\sqrt{N+1})
  \ge(2N+1)\sqrt{N}\ge(2N+1)\sqrt2.
\end{equation*}
This proves the first inequality by induction; the second uses
$t\le1$.
We also have
\begin{equation*}
  \frac9{13}<\ln 2<\frac{25}{36}.
\end{equation*}
For the lower bound, the first three terms of
$\ln 2=2\sum_{j\ge0}((2j+1)3^{2j+1})^{-1}$ sum to
$842/1215>9/13$. Replacing $2j+1$ by $3$
for every $j\ge1$ and summing the resulting geometric series gives the upper bound.
Writing $\gamma=\ln 2-1/2>5/26$, we obtain
\begin{align*}
  Nb_N&\le\frac{1}{4+t^2/2}\le\frac14-\frac{t^2}{36},\\
  18Nb_N-e_N
  &\le\frac{10}{3}+\frac{t-t^2}{2}
  \le\frac{83}{24}<\frac{45}{13}<18\gamma.
\end{align*}
Since $R_N=2e_N+2Nb_N-2\gamma$, this is equivalent to
$R_N/(2e_N)<19/18$.
For the third bound, we use the first three nonnegative terms of
the binomial expansion of $1-(1-x)^a$, where $0<a\le1$.
Let us write $s=1/N=t^2/2$ and $e=e_N$. The moment identities
\begin{equation*}
  \sum_{k=1}^N\frac{k}{2(2k-1)}=\frac N4+\frac{e_N}{2},\qquad
  \sum_{k=1}^N\frac{k^2}{2(2k-1)}=\frac{N(N+2)}8+\frac{e_N}{4}
\end{equation*}
follow by dividing the rational summands.
Together with $b_N\le1/(4N+1)$, they yield
\begin{equation}
  \label{eq:scalar-moment-expansion}
  \begin{gathered}
    \frac{C_N(u)}{2e_Nb_N}
    \ge\frac{4+s}{2}\sparen{x+\theta_2x^2+\theta_3x^3},\\
    \theta_2=\frac12-\frac s4-\frac{1}{8e},\qquad
    \theta_3=\frac13-\frac s4+\frac{s^2}{24}
    -\frac{5-2s}{48e}.
  \end{gathered}
\end{equation}
By \eqref{eq:scalar-harmonic-bound},
\begin{equation*}
  e^{-1}\le(1-t/3)^{-1}\le1+t/2.
\end{equation*}
Let
$a_2=(4+s)\theta_2/2-3/4$ and
$a_3=(4+s)\theta_3/2-1/8$.
Substitution gives
\begin{align*}
  a_2&\ge-\frac t8-\frac{5t^2}{32}-\frac{t^3}{64}
  -\frac{t^4}{32}
  \ge-\frac{21t}{64}\ge-\frac t3,\\
  a_3-\nparen{\frac13-\frac t4}
  &\ge\frac{t}{384}
  \sparen{2(1-t)(28-t)+t^2\nparen{(1-t)^2+t^3}}
  \ge0.
\end{align*}
Consequently, the difference between the right-hand side of
\eqref{eq:scalar-moment-expansion} and
$2x+3x^2/4+x^3/8$ is bounded below by
\begin{align*}
  x\sparen{\frac{t^2}{4}-\frac{tx}{3}
  +\nparen{\frac13-\frac t4}x^2}
  &=\frac{x}{12}
  \sparen{3(t-x)^2+2tx(1-x)+x^2(1-t)}\\
  &\ge0.
\end{align*}
This proves the last bound.
\end{proof}
The preceding estimates will reduce every coefficient of index at
least two to the function
\begin{equation*}
  \mathcal J(w)=\frac{55}{72}-\frac{31w^4}{8}
  +\frac{9w^8}{8}-\frac{w^{12}}8
  +38\frac{s_2(w)}w,\qquad 0<w<1.
\end{equation*}

\begin{lemma}
\label{lem:scalar-fixed-function}
For $0<w<1$, we have $\mathcal J(w)>0$.
\end{lemma}
\begin{proof}
Let us write $g(w)=s_2(w)/w$. Differentiation gives
\begin{align*}
  g'(w)&=-\frac{\ln(1-w^2)}{2w^2}-\frac12-\frac{w^2}{4}
  =\sum_{k\ge3}\frac{w^{2k-2}}{2k},\\
  g''(w)&=\frac{1}{w(1-w^2)}
  +\frac{\ln(1-w^2)}{w^3}-\frac w2.
\end{align*}
Thus
\begin{equation*}
  \frac{\mathcal J'(w)}{w^3}
  =-\frac{31}{2}+9w^4-\frac32w^8
  +19\sum_{k\ge3}\frac{w^{2k-5}}{k}
\end{equation*}
is strictly increasing on $(0,1)$.
Also $g'''(w)>0$, and for $w\ge4/5$,
\begin{equation*}
  \nparen{-\frac{31w^4}{8}+\frac{9w^8}{8}
  -\frac{w^{12}}8}'''
  =w(-93+378w^4-165w^8)>0.
\end{equation*}
Indeed, the quadratic $-93+378z-165z^2$ increases on $[0,1]$
and is positive at $z=2/5<(4/5)^4$.
Hence $\mathcal J'''(w)>0$ for $w\ge4/5$.
We need four finite estimates:
\begin{equation}
  \label{eq:scalar-fixed-checks}
  \mathcal J'(4/5)<0,\qquad
  \mathcal J''(4/5)>25,\qquad
  \mathcal J(5/6)>\frac1{256},\qquad
  |\mathcal J'(5/6)|<\frac15.
\end{equation}
For completeness, the following rational bounds certify them.
For $z>1$, let us write $v=(z-1)/(z+1)$ and define
\begin{equation*}
  P(z)=2\sum_{j=0}^5\frac{v^{2j+1}}{2j+1},\qquad
  E(z)=\frac{2v^{13}}{13(1-v^2)}.
\end{equation*}
The logarithm series gives
$P(z)<\ln z<P(z)+E(z)$.
Only $z\in\{2,3/2,5/4,11/8\}$ are needed, using
\begin{align*}
  \ln(9/25)&=2\ln(3/2)-2\ln(5/4)-2\ln 2,\\
  \ln(11/36)&=\ln(11/8)-2\ln(3/2)-\ln 2,\\
  (\ln 2)\kldivb{11/12}{1/2}&=\frac34\ln 2+\frac{11}{12}\ln(11/8)-\ln(3/2).
\end{align*}
Substituting the appropriate lower or upper bound for each logarithm
in the displayed formulas for $g',g''$, and
\begin{equation*}
  g(w)=\frac{(\ln 2)\kldivb{(1+w)/2}{1/2}-w^2/2-w^4/12}{w},
\end{equation*}
gives
\begin{equation*}
  \begin{array}{c|c|c}
    \text{quantity}&\text{strict lower bound}&\text{strict upper bound}\\ \hline
    \mathcal J'(4/5)&-1&-9/10\\
    \mathcal J''(4/5)&25&26\\
    \mathcal J(5/6)&9/2000&1/200\\
    \mathcal J'(5/6)&9/50&19/100
  \end{array}
\end{equation*}
These are finite rational comparisons and imply
\eqref{eq:scalar-fixed-checks}.
Since $\mathcal J'''$ is positive on $[4/5,1)$, we have
$\mathcal J''>25$ there. Taylor's formula at $5/6$ therefore gives
\begin{align*}
  \mathcal J(w)
  &\ge\mathcal J(5/6)+\mathcal J'(5/6)(w-5/6)
  +\frac{25}{2}(w-5/6)^2\\
  &>\frac1{256}-\frac1{1250}>0
  \qquad(4/5\le w<1).
\end{align*}
On $(0,4/5]$, the increasing ratio $\mathcal J'(w)/w^3$
and the inequality $\mathcal J'(4/5)<0$ show that
$\mathcal J$ decreases. Its value there is consequently at least
$\mathcal J(4/5)>0$.
\end{proof}
We now apply these estimates to the coefficients
\begin{equation*}
  c_N(u)=C_N(u)+\frac{4G_N(u)}u-2B_N,
  \qquad N\ge1,\quad 0<u\le1.
\end{equation*}

\begin{lemma}
\label{lem:scalar-coefficients}
For every integer $N\ge1$ and $0<u\le1$, we have $c_N(u)>0$.
\end{lemma}
\begin{proof}
For $N\ge2$ and $0<u<1$, let us write
\begin{equation*}
  k=\frac{4s_N(u)}{u b_N}-2,\qquad
  w=u^{N/2},\qquad x=1-w^4.
\end{equation*}
By \autoref{lem:scalar-tail-order},
$c_N(u)\ge C_N(u)+B_Nk$.
If $k\ge0$, this is positive because $C_N(u)>0$.
If $k<0$, \autoref{lem:scalar-moments} gives
\begin{align*}
  \frac{c_N(u)}{2e_Nb_N}
  &\ge\frac{C_N(u)}{2e_Nb_N}+\frac{R_N}{2e_N}k\\
  &\ge2x+\frac34x^2+\frac18x^3+\frac{19}{18}k.
\end{align*}
Here the upper bounds on $B_N$ and $R_N/(2e_N)$ give lower bounds
after multiplication by the negative number $k$.
Since $b_2=\ln 2-7/12<1/9$, the rescaled tail comparison gives
\begin{equation*}
  k\ge36\frac{s_2(w)}w-2.
\end{equation*}
Substituting this and $x=1-w^4$ into the preceding lower bound
gives exactly $\mathcal J(w)>0$.
For $u=1$, we instead have $s_j(1)=b_j$ and $C_N(1)=0$,
so $c_N(1)=2B_N>0$.
It remains to treat $N=1$.
Writing $\gamma=\ln 2-1/2<7/36$ and retaining the first two terms
of $s_1(u)$ gives
\begin{align*}
  c_1(u)&=\frac{1-u^2}{2}+\frac{4s_1(u)}u-2\gamma\\
  &\ge Z(u):=\frac19-\frac{u^2}{2}
  +\frac{u^3}{3}+\frac{2u^5}{15}.
\end{align*}
To see that $Z>0$, note that
$Z'(u)/u=-1+u+2u^3/3$ increases and is negative at $u=2/3$.
Also $Z''(u)=-1+2u+8u^3/3>1$ on $[2/3,1]$.
At $a=3/4$,
\begin{equation*}
  Z(a)=\frac{49}{23040}>\frac1{512},\qquad
  Z'(a)=\frac3{128}<\frac1{32}.
\end{equation*}
Taylor's formula on $[2/3,1]$ yields
\begin{equation*}
  Z(u)\ge Z(a)+Z'(a)(u-a)+\frac12(u-a)^2
  >\frac1{512}-\frac1{2048}>0.
\end{equation*}
On $(0,2/3]$, the monotonicity of $Z'(u)/u$ makes $Z$
decreasing, so $Z(u)\ge Z(2/3)>0$ there as well.
\end{proof}

\begingroup
\newtheorem*{restatedthreebit}{Lemma~\ref{lem:three-bit}}
\begin{restatedthreebit}[restated]
\begin{align*}
  &\frac{\muti{A}{B\emid C}}
  {2\left(1+\ln2\,
  \frac{\tsmuti{A}{B}-\muti{A}{B}}{\tsent{A\emid B}}\right)}
  +
  \frac{2\bigl[\tsmuti{A}{B}\muti{B}{C}-\muti{A}{C}\bigr]}
  {\sqrt{\tsmuti{B}{C}}}
  \ge\tsmuti{A}{B}-\muti{A}{B}.
\end{align*}
\end{restatedthreebit}
\endgroup
\begin{proof}
The divergence-tail definitions give the absolutely convergent expansions
\begin{align*}
  \alpha&=\sum_{N\ge1}b_Nr^N,\\
  \frac{\ln2}{1-\rho^2}
  \bigl[\rho^2 f(u)-f(\rho u)\bigr]
  &=\sum_{N\ge1}s_N(u)r^N,\\
  \frac{\ln2}{1-\rho^2}
  \bigl[f(\rho)-f(\rho u)\bigr]
  &=\sum_{N\ge1}C_N(u)r^N.
\end{align*}
In particular, $\alpha>0$.
Multiplying these series, with the convention $b_0=1$, gives
\begin{equation*}
  \begin{aligned}
    &\frac{\ln2}{1-\rho^2}\Biggl[
    f(\rho)-f(\rho u)
    +\frac{4(1+\alpha)\bigl(\rho^2 f(u)-f(\rho u)\bigr)}u\\
    &\hspace{35mm}
    -2\bigl(\rho^2-f(\rho)\bigr)(1+\alpha)\Biggr]
    =\sum_{N\ge1}c_N(u)r^N.
  \end{aligned}
\end{equation*}
Every coefficient on the right is positive by
\autoref{lem:scalar-coefficients}. Since $0<r<1$, the expression
in brackets is positive. Dividing it by $2(1+\alpha)$ proves
\eqref{eq:three-bit-scalar-target}.
\end{proof}

\section{Statement on AI use}
The proof has been obtained with interactive collaboration
with GPT-6 Astra and DeepSeek-V4.1-Flash.
\appendix
% !TeX root = courtade-kumar.tex
\section{Remaining proofs}
\label{app:remaining}
\label{app:continuous}
\label{app:comparison}

We use the setup of \autoref{sec:continuous}.

\begin{lemma}
\label{lem:ck+vscontinuous}
For the threshold bit $T_\mu$ defined in \autoref{sec:continuous},
we have
\begin{equation*}
  \qent{T_\mu}\le\muti{T_\mu}{X_\circ}.
\end{equation*}
For $0<p<1/2$, equality holds exactly when $\mu\in\{0,1/2,1\}$.
At $p=0,1/2$, equality holds for every $\mu$.
\end{lemma}
\begin{proof}
The cases $p\in\{0,1/2\}$ or $\mu\in\{0,1/2,1\}$ follow
directly. Let us fix $0<\mu<1$, $\mu\ne1/2$, and write
\begin{equation*}
  \rho=1-2p,\qquad
  F(\rho)=\muti{T_\mu}{X_\circ}-\tsent{T_\mu}\sI.
\end{equation*}
We have $F(0)=F'(0)=0$, $F(1)=\ent{T_\mu}-\tsent{T_\mu}$,
and differentiation gives
\begin{equation*}
  F''(\rho)=
  \frac{\tsent{T_\mu}(1-\tsent{T_\mu})(1-\rho)}
  {\ln2\,(1+\rho)\bigl(4\rho+\tsent{T_\mu}(1-\rho)^2\bigr)}.
\end{equation*}
This is strictly decreasing: the numerator decreases and both
factors in the denominator increase. Hence, for $0<\rho<1$,
\begin{equation*}
  \frac{F(\rho)}{\rho^2}
  =\int_0^1(1-t)F''(\rho t)\,\mathrm{d}t
  >F(1).
\end{equation*}
Since $\qI=\rho^2$, this is precisely
$\muti{T_\mu}{X_\circ}>\qent{T_\mu}$.
\end{proof}
% !TeX root = courtade-kumar.tex

We prove the properties of the comparison function used in
\autoref{sec:comparison} and the estimate used in
\autoref{lem:comparison-correlation}.
All endpoint expressions are understood by continuity.

For the following calculations, let us write
\begin{equation*}
  f(t)=\kldivb{(1+t)/2}{1/2}
  =\sum_{k\ge1}\frac{t^{2k}}{2k(2k-1)\ln2},
  \qquad -1\le t\le1.
\end{equation*}
Thus $f$ is even, convex, and strictly increasing on $[0,1]$.
Its nonnegative coefficients sum to one, so $f(t)\le t^2$.

Let $A,B$ be unbiased bits, with absolute correlation $\rho$.
Let $C$ be a nonconstant bit such that $A-B-C$ is a Markov chain,
and write
\begin{equation*}
  V=\tsent{C},\qquad
  d=\E[(2B-1)(2C-1)].
\end{equation*}
We define $F_\rho$ on $[0,1]$ by
\begin{equation*}
  F_\rho(f(t))=f(\rho t),\qquad 0\le t\le1.
\end{equation*}

\begin{lemma}
\label{lem:comparison-calculus}
The function $F_\rho$ is increasing and concave, and
\begin{equation*}
  \muti{B}{C}\le V,
  \qquad
  \muti{A}{C}\le V F_\rho\left(\frac{\muti{B}{C}}V\right).
\end{equation*}
Moreover,
\begin{equation*}
  |d|\le V,
  \qquad \muti{B}{C}\ge V f(d/V).
\end{equation*}
\end{lemma}
\begin{proof}
Let us write $x=\E[2C-1]$, so $|x|+|d|\le1$ and
$V=1-x^2$. Bayes' rule gives
\begin{align*}
  \muti{B}{C}
  &=\frac{1+x}{2}f\left(\frac{d}{1+x}\right)
  +\frac{1-x}{2}f\left(\frac{d}{1-x}\right)
  =:\ell(x,d),\\
  \muti{A}{C}&=\ell(x,\rho d).
\end{align*}
Relabeling $A$ if necessary makes its correlation with $B$
nonnegative and does not change either information.
Since $f(t)\le t^2$,
\begin{equation*}
  \ell(x,d)\le\frac{d^2}{V}\le V,
\end{equation*}
where $|d|\le1-|x|\le V$ proves the second inequality.

The divergence series also gives
\begin{equation*}
  \frac{\ell(x,d)}V
  =\sum_{k\ge1}
  \frac{W_k(x)(d/V)^{2k}}{2k(2k-1)\ln2},
  \qquad
  W_k(x)=\frac{(1+x)^{2k-1}+(1-x)^{2k-1}}2.
\end{equation*}
Here $W_1=1$ and $W_k\ge1$. This proves
$\ell(x,d)\ge Vf(d/V)$, as required for the lower bound.

Let us choose $\tau\in[0,1]$ so that $f(\tau)=\ell(x,d)/V$.
Cauchy--Schwarz gives $W_k^2\le W_{k-1}W_{k+1}$ for $k\ge2$.
Together with $W_1=1$ and $W_k\ge1$, this implies that
$W_k^{1/k}$ is nondecreasing. Indeed, successive differences of
$\ln W_k$ are nondecreasing, so
\begin{equation*}
  \ln W_{k+1}-\ln W_k\ge\frac{\ln W_k}{k-1}
  \qquad(k\ge2).
\end{equation*}
Consequently the terms
\begin{equation*}
  \frac{W_k(x)(d/V)^{2k}-\tau^{2k}}{2k(2k-1)\ln2}
\end{equation*}
change sign at most once, from negative to positive.
Their absolutely convergent sum is zero by the choice of $\tau$.
Multiplication by the decreasing weights $\rho^{2k}$ therefore
gives a nonpositive sum: subtracting a weight at the sign change
makes each summand nonpositive.
Thus
\begin{equation*}
  \ell(x,\rho d)\le Vf(\rho\tau)
  =VF_\rho\bigl(\ell(x,d)/V\bigr).
\end{equation*}
The cases with no sign change, and $\rho=0,1$, follow directly.

It remains to prove concavity. For $y=f(t)$ with $0<t<1$,
\begin{equation*}
  F_\rho'(y)=\rho^2
  \frac{\sum_{j\ge0}\rho^{2j}t^{2j}/(2j+1)}
  {\sum_{j\ge0}t^{2j}/(2j+1)}\ge0.
\end{equation*}
This ratio decreases with $t$: differentiating and pairing
terms indexed by $j,k$ gives factors
$(j-k)(\rho^{2j}-\rho^{2k})\le0$.
Since $y$ increases with $t$, the derivative decreases with $y$.
Continuity gives the assertion on $[0,1]$.
\end{proof}

Finally, the lower pointwise comparison averages in the same way.
In the setup of \autoref{sec:comparison}, let us write
$v=\tsent{C\emid U}>0$ and
$d=\E[(2B-1)(2C-1)]$.
For each value $u$ of $U$, let us write $V_u$ and $d_u$ for the
corresponding conditional entropy and correlation.
Independence of $B$ and $U$ gives $\E_{u\sim U}d_u=d$.
Since $|d_u|\le V_u$, we have $|d|\le v$.
Jensen's inequality for the convex function $f$, with weights
$V_u/v$, gives
\begin{equation*}
  \muti{B}{C\emid U}
  \ge\E_{u\sim U}\bigl[V_u f(d_u/V_u)\bigr]
  \ge v f(d/v).
\end{equation*}
Values with $V_u=0$ have $d_u=0$ and contribute zero.

% !TeX root = courtade-kumar.tex
\section{Shannon and Tsallis information}
\label{sec:quadi}

\begin{lemma}
\label{lem:binary-divergence-comparison}
For $a\in[0,1]$ and $b\in[1/2,1)$,
\begin{equation*}
  \frac{2(a-b)^2}{\ln2}
  \le\kldivb{a}{b}
  \le\frac{\ln(1/b)}{(1-b)^2\ln2}(a-b)^2.
\end{equation*}
The lower bound holds for every $b\in(0,1)$; complementing
$a,b$ gives the corresponding upper bound for $b<1/2$.
\end{lemma}
\begin{proof}
The second derivative of the divergence in $a$ is
$1/(a(1-a)\ln2)\ge4/\ln2$, and its value and first derivative
vanish at $a=b$. This proves the lower bound.
For the upper bound, Taylor's integral formula gives
\begin{equation*}
  \frac{\kldivb{a}{b}}{(a-b)^2}
  =\frac1{\ln2}\int_0^1
  \frac{1-t}{(b+t(a-b))(1-b-t(a-b))}\,\mathrm dt.
\end{equation*}
The quotient extends continuously to $a=b$ and is convex in
$a$, because $1/(x(1-x))=1/x+1/(1-x)$ is convex.
It therefore attains its maximum at an endpoint.
For $b\ge1/2$, the value at $a=1$ is at least the value at
$a=0$: equivalently,
\begin{equation*}
  -b^2\ln b+(1-b)^2\ln(1-b)\ge0.
\end{equation*}
The expression vanishes at $b=1/2$ and $b=1$, and its second
derivative is $2\ln((1-b)/b)\le0$ on this interval.
The value at $a=1$ is the claimed coefficient.
\end{proof}

For bits, we write $\tsdivb{a}{b}=4(a-b)^2$ and
$\tsent{a}=4a(1-a)$.
The binary divergence has the expansion
\begin{equation}
  \label{eq:binary-divergence-series}
  \kldivb{\tfrac{1+u}{2}}{\tfrac12}
  =\sum_{k\ge1}\frac{u^{2k}}{2k(2k-1)\ln2},
  \qquad -1\le u\le1.
\end{equation}
Indeed, expanding its second derivative
$1/((1-u^2)\ln2)$ as a geometric series and integrating twice
proves the identity for $|u|<1$; continuity gives the endpoints.
The coefficients $c_k=1/(2k(2k-1)\ln2)$ are nonnegative and
satisfy $\sum_k c_k=1$ (set $u=1$). Hence,
\begin{equation*}
  \kldivb{\tfrac{1+u}2}{\tfrac12}=\sum_{k\ge1}c_ku^{2k}
  \le u^2=\tsdivb{\tfrac{1+u}2}{\tfrac12},
\end{equation*}
with equality iff $u\in\{-1,0,1\}$. Since
$\entb{a}=1-\kldivb{a}{1/2}$ and
$\tsent{a}=1-\tsdivb{a}{1/2}$, this gives
\begin{equation*}
  \tsent{a}\le\entb{a}\quad(0\le a\le1),
  \qquad
  \muti{B}{Z}\le\tsmuti{B}{Z}\quad\text{when $B$ is a uniform bit}.
\end{equation*}
In particular $\tsent{B}\le\ent{B}$ and
$\muti{Y_1}{X}=\sI\le\qI=\tsmuti{Y_1}{X}$,
the orderings used in \autoref{fig:q-square}. The second
inequality needs the uniform reference: for a nonconstant biased
bit $B$, taking $Z=B$ gives
$\muti{B}{B}=\ent{B}>\tsent{B}=\tsmuti{B}{B}$.

\begin{proof}[Proof of \autoref{lem:tsallis-properties}]
Let us write $\mu_Z=\Pr[C=1\emid Z]$ and $\mu=\E C$.
Expanding the definition gives
\begin{equation*}
  \tsmuti{C}{Z}=4\E\bigl[(\mu_Z-\mu)^2\bigr].
\end{equation*}
This proves nonnegativity, while the upper bound by $\tsent{C}$
follows from $\tsent{C\emid Z}\ge0$. Both statements may be
averaged over $U$.
We already have $\tsent{C}\le\ent{C}$. The lower bound of
\autoref{lem:binary-divergence-comparison} with reference mean
$1/2$ also gives
\begin{equation*}
  1-\ent{C}=\kldivb{\mu}{1/2}
  \ge\frac{1-\tsent{C}}{2\ln2}
  \ge\frac{1-\tsent{C}}2.
\end{equation*}
Averaging over $U$ completes part~\ref{item:tsallis-bounds}.

The chain rule follows by telescoping the entropy differences.
If $C-Z-W$ is a Markov chain, then
$\tsent{C\emid Z,W}=\tsent{C\emid Z}$, so the chain rule
also gives the data processing inequality in
part~\ref{item:tsallis-chain}.

The lower bound of \autoref{lem:binary-divergence-comparison},
averaged over the conditional mean $\mu_Z$, gives
\begin{equation*}
  \muti{C}{Z}\ge\frac{2\E[(\mu_Z-\mu)^2]}{\ln2}
  =\frac{\tsmuti{C}{Z}}{2\ln2}.
\end{equation*}
The upper comparison for unbiased $C$ was proved above using
\eqref{eq:binary-divergence-series}. Averaging both comparisons
over $U$ gives part~\ref{item:tsallis-shannon}; the upper one
requires $C$ to remain unbiased given $U$. If a conditional law
of $C$ is constant, both informations are zero.

For two bits, evaluating the two conditional means gives
\begin{equation*}
  \tsent{C}\tsmuti{B}{C}
  =16\operatorname{Cov}(B,C)^2
  =\tsent{B}\tsmuti{C}{B}.
\end{equation*}
If either bit is constant, all three terms vanish.
When $B$ is unbiased, the definition also gives
\begin{equation*}
  \tsmuti{C}{B}
  =\bigl(\Pr[C=1\emid B=1]-\Pr[C=1\emid B=0]\bigr)^2
  =\bigl(2\Pr[B=C]-1\bigr)^2.
\end{equation*}
This proves part~\ref{item:tsallis-bits}.

For part~\ref{item:tsallis-conditioning}, let us write
$V_u=\tsent{C\emid U=u}$ and
$\sigma_u=\operatorname{Cov}(B,C\emid U=u)$.
The identity for two bits, applied conditionally, and
Cauchy--Schwarz give
\begin{align*}
  \tsmuti{B}{C\emid U}\,\tsent{C\emid U}
  &=16\left(\E_{u\sim U}\frac{\sigma_u^2}{V_u}\right)
  \left(\E_{u\sim U}V_u\right)\\
  &\ge16\left(\E_{u\sim U}\sigma_u\right)^2.
\end{align*}
Terms with $V_u=0$ have $\sigma_u=0$ and contribute zero.
Since $B$ is independent of $U$, its conditional mean is constant,
and hence $\E_{u\sim U}\sigma_u=\operatorname{Cov}(B,C)$.
The last display is therefore at least
$\tsent{B}\tsmuti{C}{B}$, as required.

Finally, applying the Shannon comparisons to the unbiased noise
bits gives $\qI/(2\ln2)\le\sI\le\qI$. Since $\ln2<1$,
we have $\qI\le2\sI$, proving part~\ref{item:tsallis-noise}.
\end{proof}

\begin{proof}[Proof of \autoref{lem:biased-comparison}]
Complementing $C$ if necessary lets us assume $\mu\ge1/2$.
Averaging the upper bound of
\autoref{lem:binary-divergence-comparison} over the conditional
mean $\Pr[C=1\mid U]$ gives the first inequality.
For the second, $\ln(1+x)\le x$ gives
\begin{equation*}
  \tsent{C}\,c(\mu)
  =\frac{\mu}{1-\mu}\ln\left(1+\frac{1-\mu}{\mu}\right)
  \le1.
\end{equation*}
\end{proof}

\printbibliography

@Article(         AhlswedeG1976,
  author        = {Ahlswede, Rudolf and Gács, Péter},
  title         = {Spreading of Sets in Product Spaces and Hypercontraction of
                  the {M}arkov Operator},
  journal       = {The Annals of Probability},
  volume        = {4},
  number        = {6},
  pages         = {925--939},
  year          = {1976}
)

@InProceedings(   AnantharamBCJN2017,
  author        = {Anantharam, Venkat and Bogdanov, Andrej and Chakrabarti, Amit
                  and Jayram, T. S. and Nair, Chandra},
  title         = {A Conjecture Regarding Optimality of the Dictator Function
                  under Hellinger Distance},
  booktitle     = {Information Theory and Applications Workshop (ITA)},
  year          = {2017}
)

@Article(         Arikan2009,
  author        = {Arikan, Erdal},
  title         = {Channel Polarization: A Method for Constructing
                  Capacity-Achieving Codes for Symmetric Binary-Input Memoryless
                  Channels},
  journal       = {IEEE Transactions on Information Theory},
  volume        = {55},
  number        = {7},
  pages         = {3051--3073},
  year          = {2009},
  doi           = {10.1109/TIT.2009.2021379}
)

@InProceedings(   BarnesO2020,
  author        = {Barnes, Leighton Pate and Özgür, Ayfer},
  title         = {The {C}ourtade-{K}umar Most Informative Boolean Function
                  Conjecture and a Symmetrized {L}i-{M}édard Conjecture are
                  Equivalent},
  booktitle     = {2020 IEEE International Symposium on Information Theory
                  (ISIT)},
  year          = {2020},
  pages         = {2205--2209},
  doi           = {10.1109/ISIT44484.2020.9174063},
  eprint        = {2004.01277},
  archiveprefix = {arXiv}
)

@Misc(            ChandarT2014,
  author        = {Chandar, Venkat and Tchamkerten, Aslan},
  title         = {Most Informative Quantization Functions},
  year          = {2014},
  howpublished  = {Manuscript},
  url           = {http://perso.telecom-paristech.fr/~tchamker/CTAT.pdf}
)

@Misc(            ChenGJL+2026,
  author        = {Chen, Zijie and Gohari, Amin and Javanmard, Adel and Lin,
                  Honghao and Mirrokni, Vahab and Nair, Chandra and Woodruff,
                  David P.},
  title         = {A Proof of the Most Informative {B}oolean Function
                  Conjecture},
  year          = {2026},
  eprint        = {2609.24931},
  archiveprefix = {arXiv},
  primaryclass  = {cs.DS}
)

@InProceedings(   ChenN2024,
  author        = {Chen, Zijie and Nair, Chandra},
  title         = {On the Optimality of Dictator Functions and Isoperimetric
                  Inequalities on Boolean Hypercubes},
  booktitle     = {2024 IEEE International Symposium on Information Theory
                  (ISIT)},
  year          = {2024},
  pages         = {3380--3385},
  doi           = {10.1109/ISIT57864.2024.10619180}
)

@Article(         CourtadeK2014,
  author        = {Courtade, Thomas A. and Kumar, Gowtham R.},
  title         = {Which Boolean Functions Maximize Mutual Information on Noisy
                  Inputs?},
  journal       = {IEEE Transactions on Information Theory},
  volume        = {60},
  number        = {8},
  pages         = {4515--4525},
  year          = {2014},
  doi           = {10.1109/TIT.2014.2326877}
)

@Article(         FischerKRSS2004,
  author        = {Fischer, Eldar and Kindler, Guy and Ron, Dana and Safra,
                  Shmuel and Samorodnitsky, Alex},
  title         = {Testing juntas},
  journal       = {Journal of Computer and System Sciences},
  volume        = {68},
  number        = {4},
  pages         = {753--787},
  year          = {2004},
  doi           = {10.1016/j.jcss.2003.11.004}
)

@Misc(            JavanmardW2026,
  author        = {Javanmard, Adel and Woodruff, David P.},
  title         = {Progress on the {C}ourtade-{K}umar Conjecture: Optimal
                  High-Noise Entropy Bounds and Generalized Coordinate-wise
                  Mutual Information},
  year          = {2026},
  eprint        = {2601.09679},
  archiveprefix = {arXiv},
  primaryclass  = {cs.IT}
)

@Misc(            KindlerOW2015,
  author        = {Kindler, Guy and O'Donnell, Ryan and Witmer, David},
  title         = {Remarks on the Most Informative Function Conjecture at fixed
                  mean},
  year          = {2015},
  eprint        = {1506.03167},
  archiveprefix = {arXiv},
  primaryclass  = {cs.IT}
)

@InProceedings(   KumarC2013,
  author        = {Kumar, Gowtham R. and Courtade, Thomas A.},
  title         = {Which Boolean Functions are Most Informative?},
  booktitle     = {2013 IEEE International Symposium on Information Theory},
  year          = {2013},
  pages         = {226--230},
  doi           = {10.1109/ISIT.2013.6620221},
  eprint        = {1302.2512},
  archiveprefix = {arXiv}
)

@Misc(            KyT2026,
  author        = {Ky, Vu Khac and Tran, Tuan},
  title         = {Dictators are most informative},
  year          = {2026},
  eprint        = {2609.24184},
  archiveprefix = {arXiv},
  primaryclass  = {cs.IT}
)

@Article(         LiM2021,
  author        = {Li, Jiange and Médard, Muriel},
  title         = {Boolean Functions: Noise Stability, Non-Interactive
                  Correlation Distillation, and Mutual Information},
  journal       = {IEEE Transactions on Information Theory},
  volume        = {67},
  number        = {2},
  pages         = {778--789},
  year          = {2021},
  doi           = {10.1109/TIT.2020.3041028},
  eprint        = {1801.04462},
  archiveprefix = {arXiv}
)

@Misc(            MahdavifarB2026,
  author        = {Mahdavifar, Hessam and Beirami, Ahmad},
  title         = {The Most Informative Bit and Beyond: A Proof of the
                  {C}ourtade--{K}umar Conjecture and Multibit Extensions},
  year          = {2026},
  eprint        = {2609.26444},
  archiveprefix = {arXiv},
  primaryclass  = {cs.IT}
)

@Misc(            OpenAI2026a,
  author        = {{OpenAI}},
  title         = {Hellinger contraction with arbitrary {B}oolean output bias},
  howpublished  = {OpenAI Math Release preprint},
  year          = {2026},
  note          = {Dated September 24, 2026; publicly released October 6, 2026},
  url           = {https://github.com/openai/math/blob/main/preprints/Hellinger-contraction-with-arbitrary-Boolean-output-bias-September-24-2026/main.pdf}
)

@Misc(            OpenAI2026b,
  author        = {{OpenAI}},
  title         = {Sharp binary-information contraction on the discrete cube},
  howpublished  = {OpenAI Math Release preprint},
  year          = {2026},
  note          = {Dated September 24, 2026; publicly released October 6, 2026},
  url           = {https://github.com/openai/math/blob/main/preprints/Sharp-binary-information-contraction-on-the-discrete-cube-September-24-2026/main.pdf}
)

@Misc(            OrdentlichSW2015,
  author        = {Ordentlich, Or and Shayevitz, Ofer and Weinstein, Omri},
  title         = {An Improved Upper Bound for the Most Informative Boolean
                  Function Conjecture},
  year          = {2015},
  eprint        = {1505.05794},
  archiveprefix = {arXiv},
  primaryclass  = {cs.IT}
)

@Article(         ParnasRS2002,
  author        = {Parnas, Michal and Ron, Dana and Samorodnitsky, Alex},
  title         = {Testing Basic Boolean Formulae},
  journal       = {SIAM Journal on Discrete Mathematics},
  volume        = {16},
  number        = {1},
  pages         = {20--46},
  year          = {2002},
  doi           = {10.1137/S0895480101407444}
)

@InProceedings(   PichlerMP2016,
  author        = {Pichler, Georg and Matz, Gerald and Piantanida, Pablo},
  title         = {A Tight Upper Bound on the Mutual Information of Two Boolean
                  Functions},
  booktitle     = {2016 IEEE Information Theory Workshop (ITW)},
  year          = {2016},
  pages         = {16--20},
  doi           = {10.1109/ITW.2016.7606787}
)

@Misc(            Samorodnitsky2015,
  author        = {Samorodnitsky, Alex},
  title         = {The "Most informative boolean function" conjecture holds for
                  high noise},
  year          = {2015},
  eprint        = {1510.08656},
  archiveprefix = {arXiv},
  primaryclass  = {cs.IT}
)

@Article(         Samorodnitsky2016,
  author        = {Samorodnitsky, Alex},
  title         = {On the Entropy of a Noisy Function},
  journal       = {IEEE Transactions on Information Theory},
  volume        = {62},
  number        = {10},
  pages         = {5446--5464},
  year          = {2016},
  doi           = {10.1109/TIT.2016.2584625}
)

@Misc(            Sarbu2017,
  author        = {Sarbu, Septimia},
  title         = {On the {C}ourtade-{K}umar conjecture for certain classes of
                  Boolean functions},
  year          = {2017},
  eprint        = {1702.03953},
  archiveprefix = {arXiv},
  primaryclass  = {cs.IT}
)

@Misc(            Sarbu2017b,
  author        = {Sarbu, Septimia},
  title         = {The maximum mutual information between the output of a
                  discrete symmetric channel and several classes of Boolean
                  functions of its input},
  year          = {2017},
  eprint        = {1701.05014},
  archiveprefix = {arXiv},
  primaryclass  = {cs.IT},
  note          = {Withdrawn}
)

@Misc(            YangW2018,
  author        = {Yang, Hengjie and Wesel, Richard D.},
  title         = {On the Most Informative Boolean Functions of the Very Noisy
                  Channel},
  year          = {2018},
  eprint        = {1807.11289},
  archiveprefix = {arXiv},
  primaryclass  = {cs.IT},
  note          = {Short version in ISIT 2019}
)

@Misc(            Yu2024,
  author        = {Yu, Lei},
  title         = {Local Optimality of Dictator Functions with Applications to
                  {C}ourtade--{K}umar and {L}i--{M}édard Conjectures},
  year          = {2024},
  eprint        = {2410.10147},
  archiveprefix = {arXiv},
  primaryclass  = {math.PR},
  note          = {Accepted for publication in the Annals of Applied
                  Probability}
)
\end{document}